\documentclass[11pt]{article}

\usepackage[margin=2cm]{geometry}
\usepackage{graphicx,amsmath,booktabs,hyperref}
\usepackage[super,comma,sort&compress]{natbib}
\usepackage{xcolor}
\usepackage{subcaption}
\usepackage{setspace}
\usepackage{authblk}
\usepackage{multirow}

\hypersetup{
    colorlinks=false,
    linkcolor=black,
    citecolor=black,
    urlcolor=black
}

\title{\textbf{The Geometry of Forgetting}}

\author[1]{Sambartha Ray Barman}
\author[1]{Andrey Starenky}
\author[1]{Sophia Bodnar}
\author[1]{Nikhil Narasimhan}
\author[1,2]{Ashwin Gopinath\thanks{Corresponding author: agopi@mit.edu, ashwin@sentra.app}}
\affil[1]{Sentra, 235 2nd Street, San Francisco, CA 94105, USA}
\affil[2]{Department of Mechanical Engineering, Massachusetts Institute of Technology, 77 Massachusetts Avenue, Cambridge, MA 02139, USA}

\date{}

\begin{document}

\maketitle


\begin{abstract}
\noindent
Why do we forget? Why do we remember things that never happened? The conventional answer points to biological hardware. We propose a different one: geometry. Here we show that high-dimensional embedding spaces, subjected to noise, interference, and temporal degradation, reproduce quantitative signatures of human memory with no phenomenon-specific engineering. Power-law forgetting ($b = 0.460 \pm 0.183$, human $b \approx 0.5$) arises from interference among competing memories, not from decay. The identical decay function without competitors yields $b \approx 0.009$, fifty times smaller. Time alone does not produce forgetting in this system. Competition does. Production embedding models (nominally 384--1{,}024 dimensions) concentrate their variance in only ${\sim}16$ effective dimensions, placing them deep in the interference-vulnerable regime. False memories require no engineering at all: cosine similarity on unmodified pre-trained embeddings reproduces the Deese--Roediger--McDermott false alarm rate ($0.583$ versus human ${\sim}0.55$) with zero parameter tuning and no boundary conditions. We did not build a false memory system. We found one already present in the raw geometry of semantic space. These results suggest that core memory phenomena are not bugs of biological implementation but features of any system that organizes information by meaning and retrieves it by proximity.
\end{abstract}


Every student of psychology learns the same story: human memory is powerful but broken. We forget on a power-law curve first documented by Ebbinghaus\cite{ebbinghaus1885} and confirmed across dozens of paradigms\cite{wixted1991}. We confidently ``remember'' events that never occurred\cite{roediger1995}. We fail to retrieve answers we demonstrably know\cite{brown1966}. Distributed practice beats massed repetition\cite{cepeda2006}, consolidation reorganizes stored representations\cite{nadel1997,mcclelland1995}, and cross-modal binding links visual, auditory, and linguistic traces into coherent episodes\cite{baddeley2000,tulving1972}. The standard explanation blames biology: the brain does its best, but evolution left us with hardware that leaks. This paper offers a different explanation. We show that these ``flaws'' arise from the mathematics of similarity-based retrieval in high-dimensional space, and they appear in systems that have nothing to do with biology. Geometry, not wetware, may be the deeper cause. Biology determines the parameter regime; geometry determines the failure modes. If this is right, then much of what we call ``forgetting'' and ``false memory'' was never a bug. It is what any system that organizes information by meaning is bound to do.

What makes us forget? Two theories have competed for over a century. Decay theories say memory traces fade with time. Interference theories say they are crowded out by competitors\cite{bjork1992,anderson1991}. Our experiments address one specific mechanism within this debate: retrieval competition among stored embeddings, not the full cognitive story. The new theory of disuse\cite{bjork1992} attempts to reconcile these positions by distinguishing retrieval strength (current accessibility, which declines with competing associations) from storage strength (relatively permanent). Under this framework, forgetting reflects a decline in retrieval strength caused by interference, not the dissolution of the trace itself. Empirical evidence increasingly favours interference: forgetting is accelerated by subsequently learned material\cite{mccloskey1989} and modulated by the similarity between competing items, as predicted by competition but not by passive decay. Yet why certain representational architectures are vulnerable to interference and others are not has remained an open question.

Modern embedding models offer an unexpected window into this question. Transformer-based encoders\cite{vaswani2017} map sentences into high-dimensional spaces where meaning is captured by angle. Sentence-level models\cite{reimers2019} extend this to variable-length text; contrastive models like CLIP\cite{radford2021} span sensory modalities; dense retrievers\cite{xiao2023} and large language models\cite{qwen2024} push semantic structure further. None of these systems were built to model memory. They were built for representation and retrieval. But the geometric properties they exhibit (semantic clustering, concentration of measure, sensitivity of angular proximity to perturbation) are the properties that interference theories predict should govern forgetting. The present experiments focus on embedding-based retrieval spaces. Representational encoders, dense retrievers, and generative models differ in important ways, but they share the same underlying geometry. The question is whether that geometry alone is enough to produce the failure modes of human memory.

\begin{figure}[!ht]
    \centering
    \includegraphics[width=\textwidth]{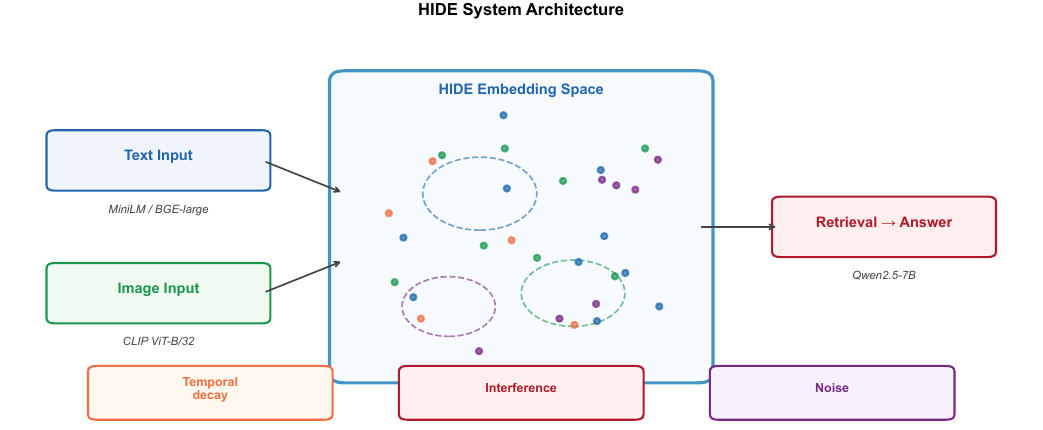}
    \caption{\textbf{Extended Data Fig.~1: System architecture.} Conceptual overview of the embedding-based memory framework. Input modalities are encoded by frozen pre-trained models, projected into a shared space, and retrieved via cosine similarity with temporal decay modulation.}
    \label{fig:architecture}
\end{figure}

It is. Across five experimental domains, using only open-weight models and public datasets (Fig.~\ref{fig:architecture}), we show that interference from competing memories, not temporal decay, is the dominant driver of power-law forgetting in the tested retrieval geometry; that vulnerability to interference is set by effective rather than nominal dimensionality; and that false memories, spacing effects, and tip-of-tongue states emerge from the geometry of pre-trained embedding spaces with no phenomenon-specific engineering. We store memories as contextually enriched embeddings (incorporating positional, temporal, and episodic metadata) and retrieve them via cosine similarity. The boundary conditions (noise, competing memories, temporal degradation) correspond to well-established features of any memory system, biological or artificial. The quantitative patterns follow from geometry.

\section*{Results}

\begin{figure}[!ht]
    \centering
    \includegraphics[width=\textwidth]{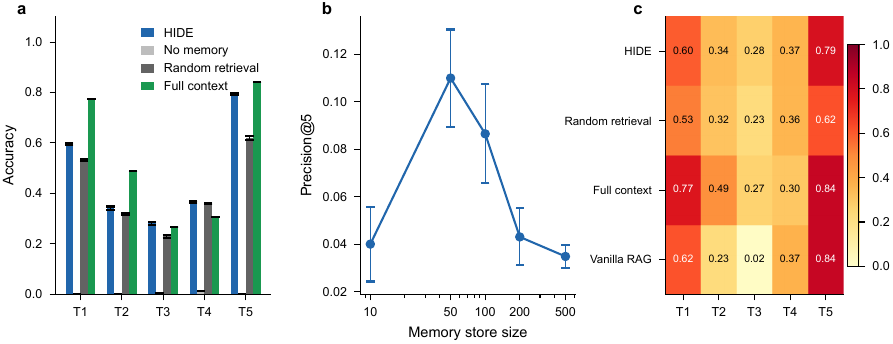}
    \caption{\textbf{Extended Data Fig.~2: Memory-dependent reasoning on bAbI.} \textbf{a},~Accuracy by task for contextual retrieval versus baselines. \textbf{b},~Memory scaling: precision declines with store size. \textbf{c},~Retrieval precision heatmap. Mean $\pm$ 95\% CI, $n = 5$ seeds.}
    \label{fig:ed_phase1}
\end{figure}

As a proof-of-concept sanity check, we first tested whether contextually enriched embeddings can function as a basic memory substrate on five memory-dependent reasoning tasks (bAbI, tasks 1--5). Mean accuracy was $0.475 \pm 0.007$ (95\% CI: $[0.470, 0.482]$), exceeding no-memory ($0.003$) and random retrieval ($0.411 \pm 0.004$) baselines on all tasks, though remaining below the full-context ceiling on most tasks (Extended Data Fig.~\ref{fig:ed_phase1}; Supplementary Table~\ref{tab:phase1_full}). Per-task gains were uneven: Task~5 showed the largest improvement ($+0.173$ over random) while Task~4 showed only marginal gain ($+0.005$). Retrieval precision declined with increasing memory load, consistent with the fan effect in human memory\cite{murdock1962}, foreshadowing the role of interference examined below.

\subsection*{Interference, not decay, produces human-like forgetting curves}

We next asked whether embedding-based memory exhibits the power-law forgetting ubiquitous in human memory. We augmented embeddings with multi-scale temporal encoding (64-dim, three sinusoidal scales) and modulated retrieval by temporal decay: $\text{score} = \cos(\mathbf{q}, \mathbf{m}) \times (1 + \beta t)^{-\psi}$, with $\psi = 0.5$\cite{wixted1991}. We simulated the classical Ebbinghaus paradigm by encoding 1{,}000 facts spanning 30 simulated days, querying at day~30, and binning retrieval accuracy by age.

\begin{figure}[!ht]
    \centering
    \includegraphics[width=\textwidth]{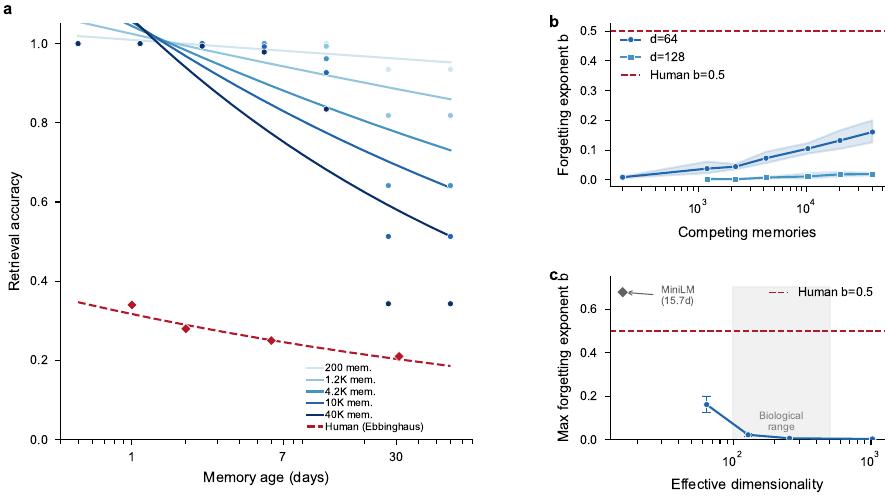}
    \caption{\textbf{Interference from competing memories produces human-like forgetting in low-dimensional embedding spaces.} \textbf{a},~Forgetting curves steepen progressively as competing memories are added (light to dark blue), converging toward the classical human forgetting curve (red dashed; Ebbinghaus, 1885). All curves at $d = 64$ with identical decay function. \textbf{b},~The fitted forgetting exponent $b$ increases monotonically with competitor count at $d = 64$ (blue) but remains near zero at $d = 128$ (grey). Human $b \approx 0.5$ (red dashed). Shaded regions: bootstrap 95\% CI. \textbf{c},~Maximum forgetting exponent as a function of effective dimensionality. Interference is substantial only below $d \approx 100$, placing biological neural codes (estimated $d = 100$--$500$\cite{stringer2019,gao2017}; shaded) near the transition zone. Orange diamond: MiniLM ($d_\text{nom} = 384$, $d_\text{eff} \approx 16$) shows strong interference consistent with its low effective dimensionality. $n = 5$ seeds throughout.}
    \label{fig:interference}
\end{figure}

The critical test: does the forgetting exponent depend on the decay function or on the number of competing memories? With decay alone and no competitors, $b \approx 0.009$, fifty times smaller than the human value. Decay by itself does not produce human-like forgetting. Adding 10{,}000 distractors while keeping the decay function unchanged raised the exponent to $b = 0.460 \pm 0.183$ (95\% CI: $[0.354, 0.644]$, $R^2 = 0.757 \pm 0.058$), within the range of human data. Time is not what produces forgetting here. Crowding is. The forgetting curves fan out progressively as competitors accumulate (Fig.~\ref{fig:interference}a), converging toward the classical Ebbinghaus curve. The dose--response relationship (Fig.~\ref{fig:interference}b) confirms that interference drives the exponent monotonically\cite{bjork1992,anderson1991}.

To dissect the geometry of interference, we varied the semantic relationship between targets and competitors and the effective dimensionality of the space. At $d = 64$, same-article (``near'') competitors produced stronger interference ($b = 0.161$, CI: $[0.126, 0.200]$ at 40{,}000 competitors) than cross-article (``far'') competitors ($b = 0.132$, CI: $[0.105, 0.156]$ at 50{,}000), confirming that confusability depends on semantic proximity. The dimensionality dependence was clear: at $d = 128$ the maximum exponent dropped to $b = 0.020$, and at $d \geq 256$ it remained below $0.004$ regardless of competitor count (Fig.~\ref{fig:interference}c). This protection arises from the concentration of measure: in higher dimensions, the probability that a competitor falls within a target's confusable neighbourhood decreases sharply. As a plausibility argument (not a direct bridge), we note that cortical representations are estimated to operate at effective dimensionalities of 100--500 from neural recordings\cite{stringer2019,gao2017}, though these estimates depend heavily on task, brain area, analysis method, and recording regime. If these estimates are in the right range, biological memory would occupy a regime where interference is non-negligible but not catastrophic, consistent with, though not proof of, a geometric contribution to interference vulnerability.

\begin{figure}[!ht]
    \centering
    \includegraphics[width=\textwidth]{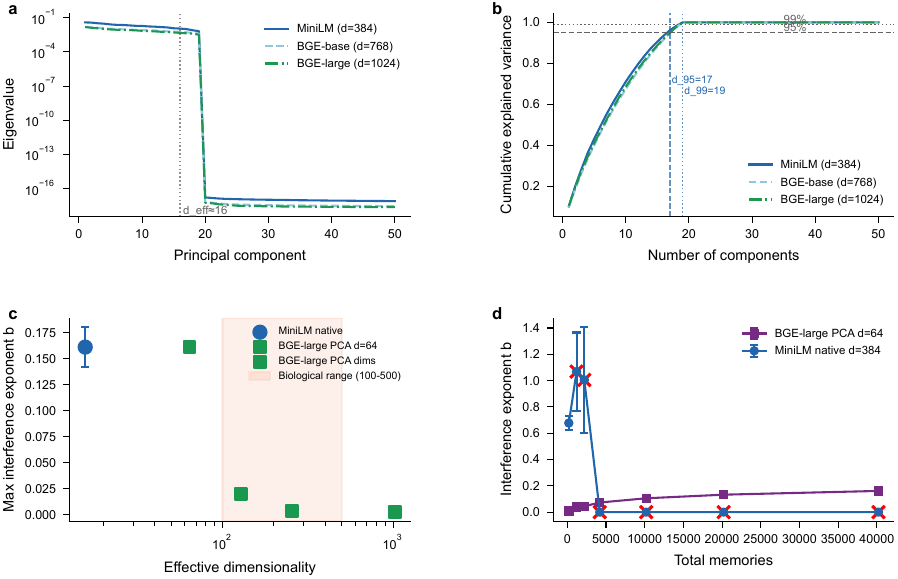}
    \caption{\textbf{Extended Data Fig.~11: Effective dimensionality resolves the dimensionality paradox.} \textbf{a},~Eigenvalue spectra of three embedding models show rapid drop-off; vertical dashed lines mark the participation ratio ($d_\text{eff}$). \textbf{b},~Cumulative explained variance: only 17--18 components account for 95\% of variance regardless of nominal dimensionality (384, 768, or 1{,}024). \textbf{c},~Maximum interference exponent $b$ as a function of effective dimensionality. MiniLM at native $d = 384$ ($d_\text{eff} \approx 16$, orange diamond) shows stronger interference than BGE-large at PCA $d = 64$, consistent with its low effective dimensionality. \textbf{d},~Direct comparison: MiniLM $d = 384$ without PCA shows catastrophic retrieval failure with near distractors, exceeding BGE-large PCA $d = 64$ interference. $n = 5$ seeds; error bars: bootstrap 95\% CI.}
    \label{fig:ed_dimensionality}
\end{figure}

\vspace{0.3em}
\noindent\textbf{The dimensionality illusion: tested embedding models fall in the interference-vulnerable regime.}\quad An apparent paradox arises: if interference requires $d \leq 64$, how can it be relevant to production embedding models with nominal dimensionality 384--1{,}024? Spectral analysis exposes what we term the \textit{dimensionality illusion} (Extended Data Fig.~\ref{fig:ed_dimensionality}). Computing the participation ratio ($d_\text{eff} = (\sum \lambda_i)^2 / \sum \lambda_i^2$) across three models reveals that all concentrate their variance in few dimensions: MiniLM-L6-v2 ($d_\text{nom} = 384$) has $d_\text{eff} = 15.7 \pm 0.0$; BGE-base ($d_\text{nom} = 768$) has $d_\text{eff} = 16.6 \pm 0.1$; BGE-large ($d_\text{nom} = 1{,}024$) has $d_\text{eff} = 16.3 \pm 0.1$. Only 17--18 principal components are needed for 95\% explained variance regardless of nominal dimensionality. To confirm the functional consequence, we ran the interference protocol on MiniLM at its native 384 dimensions without PCA projection: same-article distractors caused complete retrieval collapse at $\geq$20 distractors per target, and even without near competitors the fitted exponent reached $b = 0.678$ (CI: $[0.583, 0.789]$), far exceeding the $b = 0.161$ observed for BGE-large at PCA $d = 64$ (Fig.~\ref{fig:interference}c). The effective dimensionality, not the nominal dimensionality, determines interference vulnerability in this setup. Retrieval systems built on these embedding models may be operating in the interference-vulnerable regime regardless of their advertised dimensionality, though hybrid systems with lexical filters, metadata constraints, or rerankers may behave differently. When a model claims to be 1{,}024-dimensional but concentrates its variance in 16, the label is a misnomer, and the interference risk is real.

\subsection*{False memories emerge naturally from semantic clustering}

The strongest version of our claim is not that we can engineer human-like memory errors, but that they already exist in the raw geometry before any intervention. The Deese--Roediger--McDermott (DRM) paradigm\cite{roediger1995} is the gold standard for studying false memory: participants study word lists (e.g., \emph{bed, rest, awake, tired, dream}\ldots) and subsequently ``remember'' an unstudied but semantically associated critical lure (\emph{sleep}) at rates of ${\sim}55\%$. The conventional interpretation treats this as a failure of source monitoring or associative activation. The geometric perspective reframes it: any retrieval system that organizes items by meaning will place semantically related concepts in nearby regions, and any threshold-based decision will confuse items within those regions.

\begin{figure}[!ht]
    \centering
    \includegraphics[width=\textwidth]{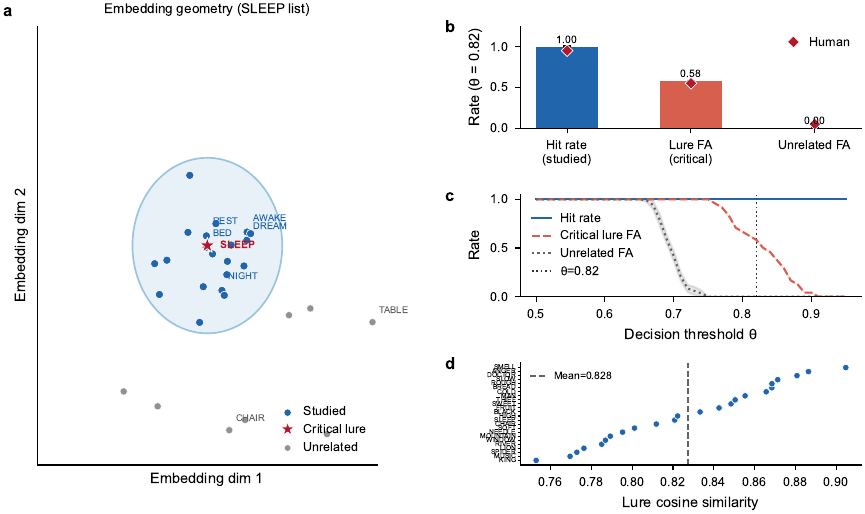}
    \caption{\textbf{False memories arise from semantic clustering in the embedding space.} \textbf{a},~UMAP projection of the SLEEP word list shows the critical lure (\emph{sleep}, red star) falling within the cluster of studied words (blue dots), while unrelated words (grey) remain separated. \textbf{b},~At $\theta = 0.82$, the critical-lure false alarm rate ($0.583$) closely matches the human value (${\sim}0.55$, red diamonds). \textbf{c},~Threshold operating curves show the lure false alarm rate remains elevated above unrelated across a wide range of thresholds. \textbf{d},~Per-list lure cosine similarity is consistently high across all 24 DRM word lists. $n = 5$ seeds for panels \textbf{b}--\textbf{d}.}
    \label{fig:drm}
\end{figure}

We replicated this paradigm using all 24 published DRM word lists encoded with a 1024-dimensional retrieval model\cite{xiao2023}. The geometry of the embedding space reveals why false memories arise naturally in this setting: critical lures occupy positions geometrically indistinguishable from studied items, falling within the cluster of semantically related words, while unrelated words remain clearly separated (Fig.~\ref{fig:drm}a). At the threshold ($\theta = 0.82$) that produces zero unrelated false alarms (an independent criterion), the critical-lure false alarm rate was $0.583$, within 3.3 percentage points of the human value (Fig.~\ref{fig:drm}b). The full threshold operating curve (Fig.~\ref{fig:drm}c) is more informative than any single operating point: the lure false alarm rate remains elevated above unrelated items across a wide range of thresholds. This result held consistently across all 24 lists (Fig.~\ref{fig:drm}d). No parameter was tuned to produce this correspondence. The phenomenon is unbaked: raw cosine similarity on raw pre-trained embeddings reproduces the effect because the geometry of semantic space naturally produces it. We note that the word lists were embedded with a sentence-level model; whether alternative encoders calibrated differently for isolated words would yield different rates remains to be tested. This result has an important asymmetry with the forgetting results: it requires no boundary conditions at all. Forgetting requires competitors. Spacing effects require noise and competitors. False memories require nothing. They sit in the geometry of meaning itself, waiting to be retrieved. The same semantic structure that makes these models useful for retrieval is the structure that produces false memories. A system that eliminates false memories of this kind would likely sacrifice the semantic structure that makes it valuable, though hybrid systems with metadata constraints or explicit verification may attenuate such errors while preserving substantial semantic utility.

\subsection*{Spaced repetition survives age-dependent degradation}

The spacing effect, where distributed practice produces stronger retention than massed repetition\cite{cepeda2006}, is one of the most robust phenomena in memory research. The mechanism behind the spacing effect, in geometric terms, is straightforward: a more recently encoded trace is less degraded by noise at test. Long-spaced practice ensures that one repetition is always relatively recent; massed practice ensures all repetitions are equally old. We note that this does not yet isolate a deep geometric principle distinct from ``latest trace dominates under age-dependent corruption.'' The spacing effect in this setup is a boundary-condition-dependent phenomenon, unlike the DRM result. We encoded 100 facts with three repetitions each under four spacing conditions and tested at 30 simulated days, introducing Wikipedia distractors and dimension-normalized age-proportional noise (see Methods).

\begin{figure}[!ht]
    \centering
    \includegraphics[width=\textwidth]{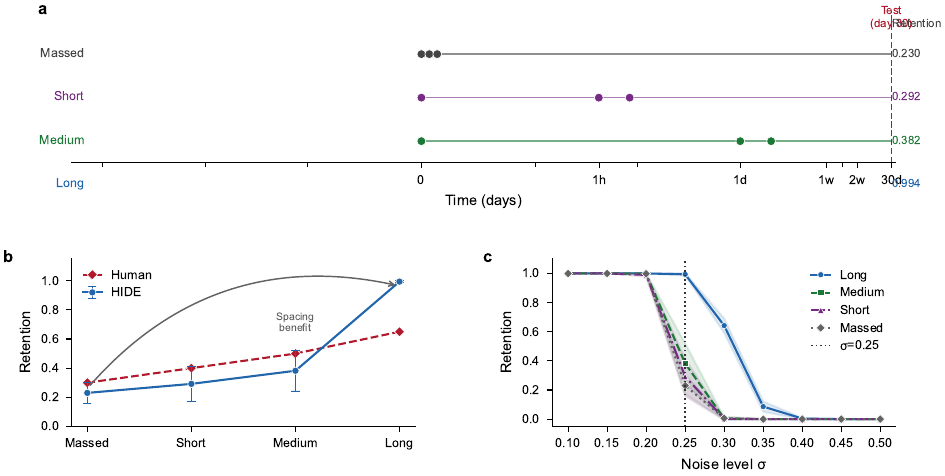}
    \caption{\textbf{Spaced repetition survives age-dependent degradation through recency of the most recent trace.} \textbf{a},~Timeline showing repetition schedules: long-spaced items retain a recent trace at test (14 days old), while massed items have only old traces (30 days). Retention values shown at right. \textbf{b},~Retention increases monotonically with spacing interval (blue), matching the human ordering (red diamonds): long $>$ medium $>$ short $>$ massed. \textbf{c},~The spacing gradient emerges as noise increases: at $\sigma = 0$, all conditions achieve ceiling; as noise grows, conditions separate, with massed dropping fastest. Selected $\sigma = 0.25$ (dashed line). Error bars: bootstrap 95\% CI, $n = 5$ seeds.}
    \label{fig:spacing}
\end{figure}

A systematic sweep confirmed this: at $\sigma = 0.25$ with 25{,}000 distractors, retention followed the expected ordering: long-spaced retention $= 0.994 \pm 0.008$, medium $= 0.382 \pm 0.139$, short $= 0.292 \pm 0.121$, and massed $= 0.230 \pm 0.073$ (Fig.~\ref{fig:spacing}b), matching the human pattern (long $>$ medium $>$ short $>$ massed, Cohen's $d = 13.1$). The mechanism is visible in the timeline (Fig.~\ref{fig:spacing}a): the most recent repetition of a long-spaced item is younger and therefore less degraded at test. The spacing gradient emerges as noise increases from ceiling performance (Fig.~\ref{fig:spacing}c), paralleling encoding variability theory\cite{cepeda2006}. The dependence on noise means that spacing here is a boundary-condition-dependent phenomenon, in contrast to the DRM result which requires no boundary conditions.

We also observed tip-of-tongue states, cases where the correct memory ranks 2--20 with high similarity, at $3.66 \pm 0.13\%$ (human ${\sim}1.5\%$)\cite{brown1966}. The qualitative phenomenon of partial retrieval emerges from retrieval competition (Extended Data Fig.~\ref{fig:ed_tot}), though the rate exceeds the human baseline by a factor of ${\sim}2.4$, suggesting that the geometric definition (correct item ranked 2--20) may be looser than the phenomenological criterion in humans. We present this as qualitative emergence, not quantitative correspondence.

\subsection*{Exploratory: topological structure of the memory manifold}

\begin{figure}[!ht]
    \centering
    \includegraphics[width=\textwidth]{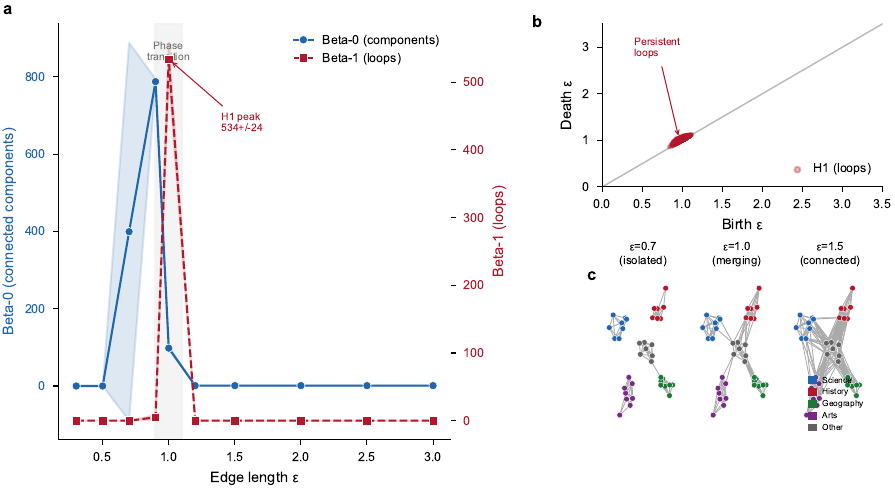}
    \caption{\textbf{Exploratory: topological structure of the memory manifold shows a connectivity transition with transient loop structure.} \textbf{a},~Connected components ($H_0$, blue) exhibit a sharp $1{,}000 \to 1$ transition while loops ($H_1$, red) peak at $534 \pm 24$ at $\epsilon = 1.0$ (shaded: phase transition zone). \textbf{b},~Persistence diagram showing long-lived topological features (far from diagonal) versus noise (near diagonal). \textbf{c},~UMAP visualizations at three edge lengths: disconnected clusters ($\epsilon = 0.7$), critical connectivity with loops ($\epsilon = 1.0$), and fully connected ($\epsilon = 1.5$). Points colored by topic. Error bands: 95\% CI, $n = 5$ seeds.}
    \label{fig:topology}
\end{figure}

As an exploratory analysis, we examined the topological structure of the memory manifold. We computed persistent homology (Rips complex)\cite{gudhi2014} on 1{,}000-point subsamples of 10{,}000 Wikipedia sentence embeddings\cite{xiao2023}. The zeroth Betti number ($H_0$, connected components) exhibited a sharp phase transition: all points were isolated at $\epsilon < 0.7$, rapid clustering began at $\epsilon = 0.9$ ($H_0 = 788 \pm 7$), and the space collapsed to a single component at $\epsilon \geq 1.2$ (Fig.~\ref{fig:topology}a). The first Betti number ($H_1$, loops) peaked at $\epsilon = 1.0$ ($H_1 = 534 \pm 24$), revealing rich non-trivial topological structure precisely at the connectivity transition. These transient loops, semantic ``holes'' that appear as the space transitions from disconnected clusters to full connectivity (Fig.~\ref{fig:topology}c), are consistent with hierarchical thematic structure in natural language. Whether and how this topological structure is mechanistically necessary for the memory phenomena under study remains an open question; we present it as descriptive characterisation rather than causal evidence.

\subsection*{Shared embedding geometry enables cross-modal retrieval}

\begin{figure}[!ht]
    \centering
    \includegraphics[width=\textwidth]{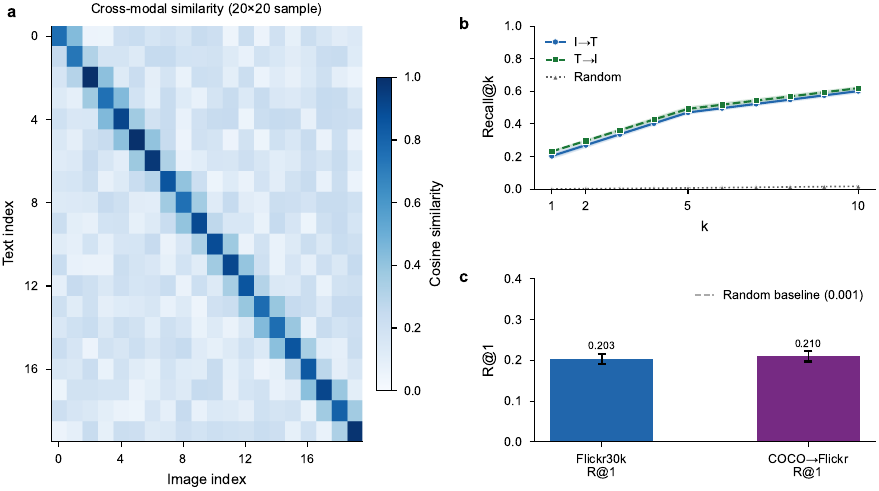}
    \caption{\textbf{Lightweight projections align independent encoders for cross-modal retrieval.} \textbf{a},~Cross-modal similarity matrix shows strong diagonal structure: image queries retrieve their matched text descriptions with high cosine similarity (dark diagonal), while unrelated pairs show low similarity. \textbf{b},~Recall at $k \in \{1, 5, 10\}$ for image-to-text and text-to-image retrieval, exceeding random baseline by two orders of magnitude. \textbf{c},~Cross-dataset transfer (COCO$\to$Flickr30k) shows minimal degradation. Error bars: 95\% CI, $n = 5$ seeds.}
    \label{fig:crossmodal}
\end{figure}

\begin{figure}[!ht]
    \centering
    \includegraphics[width=\textwidth]{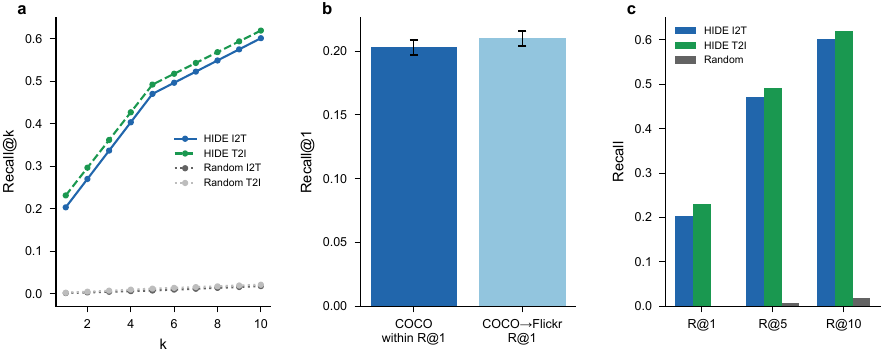}
    \caption{\textbf{Extended Data Fig.~6: Cross-modal retrieval detailed results.} \textbf{a},~Recall curves from $k = 1$ to $k = 10$ for both retrieval directions. \textbf{b},~Transfer comparison. \textbf{c},~Comparison with random baseline. $n = 5$ seeds.}
    \label{fig:ed_crossmodal}
\end{figure}

As an additional exploratory direction, we asked whether lightweight geometric alignment can achieve cross-modal binding, a feature of human episodic memory\cite{baddeley2000,tulving1972}, without dedicated biological machinery. Lightweight projection layers (${\sim}10$K parameters each) aligning independently pre-trained text\cite{reimers2019} and image\cite{radford2021} encoders into a shared space yielded image-to-text Recall@1 $= 0.203 \pm 0.013$ and text-to-image R@1 $= 0.231 \pm 0.001$ on Flickr30k, exceeding random projection by two orders of magnitude (Fig.~\ref{fig:crossmodal}). Projections transferred across datasets with minimal degradation (COCO$\to$Flickr30k R@1 $= 0.210 \pm 0.013$; Extended Data Fig.~\ref{fig:ed_crossmodal}). This result does not materially strengthen the central claim about forgetting and false memory, but it is consistent with the broader hypothesis that geometric structure in embedding spaces can support memory-like capabilities. Full detail is in Extended Data.

\subsection*{A shared mathematical substrate for biological and artificial memory}

\begin{figure}[!ht]
    \centering
    \includegraphics[width=0.5\textwidth]{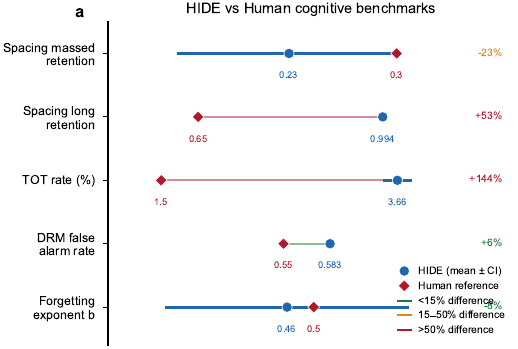}
    \caption{\textbf{Qualitative comparison to canonical human benchmarks.} For each phenomenon, the observed value (blue dot, with 95\% CI) is compared to the corresponding human reference (red diamond). Short connecting lines indicate close matches; longer lines indicate discrepancies. The forgetting exponent and DRM false alarm rate show the closest correspondence; other phenomena show qualitative but not tight quantitative agreement.}
    \label{fig:summary}
\end{figure}

Five experiments, one framework, and a pattern that is hard to dismiss as coincidence (Fig.~\ref{fig:summary}). The Ebbinghaus forgetting exponent ($b = 0.460 \pm 0.183$ with 10{,}000 distractors, human $b \approx 0.500$) and DRM false alarm rate ($0.583$, human ${\sim}0.55$) show close correspondence. The interference experiment isolates the exponent's dependence on competitor count ($b = 0.161$ at 40{,}000 near competitors, $d = 64$). The spacing effect reproduces the correct ordering with a wider dynamic range than human data. The TOT rate ($3.66\%$) exceeds the human baseline ($1.5\%$) by a factor of ${\sim}2.4$ (a qualitative but not quantitative match), and the spacing long-retention ($0.994$) overshoots the human value (${\sim}0.65$). These matches span a continuum from fully emergent (DRM, requiring no boundary conditions) to boundary-condition-dependent (spacing, requiring specific noise and distractor parameters). That these matches arise from a single framework, and that the mismatches are systematic rather than random, is itself the finding.

\section*{Discussion}

These results are not an analogy. Biological and artificial memory systems share failure modes because both are subject to the same geometric constraints: low effective dimensionality, semantic clustering, noise, and competition. The central finding is that interference from competing memories, rather than passive trace decay, is the dominant driver of power-law forgetting in the tested retrieval geometry. This aligns with and extends the Bjork \& Bjork new theory of disuse\cite{bjork1992}, providing a concrete geometric mechanism for the distinction between retrieval strength and storage strength.

The interference experiment reveals the specific geometric conditions under which forgetting occurs. In the simulation, stored embeddings retain their nominal fidelity (though the query corruption and temporal weighting jointly define effective accessibility), but the probability of retrieving the correct memory at rank~1 decreases as more competitors populate its neighbourhood, reducing retrieval strength. This is the pattern predicted by interference theory\cite{anderson1991}. The semantic proximity requirement reflects the geometric fact that confusability depends on angular distance: only items projecting into similar regions compete for retrieval. The dimensionality dependence ($d = 64$ produces measurable interference; $d \geq 128$ is effectively immune) arises from the concentration of measure: in higher dimensions, random points are exponentially unlikely to fall within any fixed angular neighbourhood. The distinction between storage and retrieval strength thus emerges naturally from cosine similarity search in a crowded, low-dimensional space.

The dimensionality result cuts both ways. Neural population codes in cortical areas operate at effective dimensionalities of 100--500\cite{stringer2019,gao2017}, placing biological memory in the regime where interference is non-negligible but not catastrophic. Human forgetting may not be a design flaw. It may be the price of admission for the computational benefits that moderate-dimensional codes provide.

\begin{figure}[!ht]
    \centering
    \includegraphics[width=\textwidth]{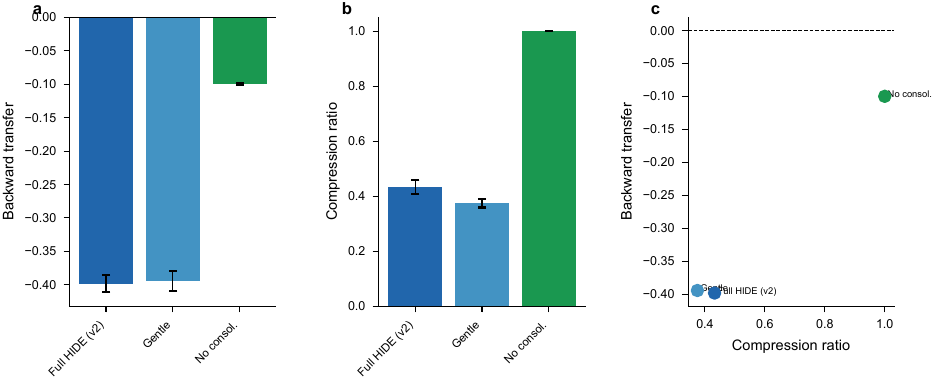}
    \caption{\textbf{Extended Data Fig.~5: Consolidation as an informative negative result.} \textbf{a},~Backward transfer worsens with consolidation despite \textbf{b},~substantial memory compression. \textbf{c},~The compression--accuracy trade-off: geometric merging achieves compression at the cost of retrieval accuracy.}
    \label{fig:ed_consolidation}
\end{figure}

For artificial systems, the implications are immediate. Production embedding models, despite nominal dimensionalities of 384--1{,}024, concentrate their variance in approximately 16 effective dimensions (Supplementary Table~\ref{tab:eff_dim}). RAG systems and long-term agent memories built on these models are likely operating in the interference-vulnerable regime, though hybrid systems with lexical filters, metadata constraints, or guardrails may behave differently. Retrieval accuracy in these systems will degrade as a power law with database size. This is not a worst-case scenario; it is the expected behaviour, predictable from first principles. Every vector database will eventually forget. The concentration-of-measure protection at high effective dimensionality suggests a clear design target: increasing the effective rank of stored representations, a target that current embedding models, optimised for semantic clustering, work directly against. The consolidation failure (Extended Data Fig.~\ref{fig:ed_consolidation}) further exposes what we term the \textit{vector averaging fallacy}: the widespread engineering practice of compressing retrieval databases or summarizing conversation histories by averaging nearby embeddings is not merely suboptimal but geometrically destructive, collapsing the angular distinctions on which similarity-based retrieval depends.

The DRM false memory result is particularly compelling because it is unbaked. Raw cosine similarity on a pre-trained embedding space, applied to published word lists without modification, produces a critical-lure false alarm rate within 3.3 percentage points of the human value. The semantic geometry simply clusters related concepts, and retrieval based on similarity naturally ``remembers'' unstudied items within the cluster. This suggests that human false memories arise from an analogous geometric mechanism: neural representations of related concepts occupy nearby regions, and pattern-completion-based retrieval confuses items within these regions.

If the same geometric mechanism underlies human false memories, the implication is uncomfortable: false memories are not errors introduced by faulty hardware. They are a cost of the same geometry that supports generalisation and pattern completion. A memory system that never confuses related concepts is a memory system that cannot generalise across them. This is a tradeoff frontier, not a strict impossibility: reducing DRM-type false memories likely requires sacrificing some of the semantic clustering that gives the system its power, though metadata constraints, source attribution, or explicit verification may attenuate false recall while preserving substantial semantic utility.

We propose a general principle for interpreting these correspondences. The phenomena span a continuum from fully emergent to boundary-condition-dependent. The DRM result requires no boundary conditions; it emerges from raw geometry. The Ebbinghaus match requires competitor memories, but the decay function alone is insufficient ($b \approx 0.009$ without competitors). The spacing effect requires both noise and competitors, with specific parameters ($\sigma = 0.25$, 25{,}000 distractors) determining the gradient. In each case, the boundary conditions correspond to well-established features of biological memory (neural noise, competing memories, temporal degradation), and the quantitative patterns emerge from geometry operating under these constraints. This hierarchy has a natural interpretation: phenomena closer to the fully emergent end are more fundamental: they will appear in any similarity-based retrieval system regardless of implementation details. Phenomena closer to the boundary-condition-dependent end are more contingent, requiring specific noise regimes and competitor densities that may differ across biological and artificial systems. The DRM result sits at the fundamental end. Forgetting sits in the middle. The spacing effect sits toward the contingent end. The observation that all three emerge from the same framework, at different points on this continuum, is itself evidence for a unified geometric account rather than a collection of independent mechanisms.

The consolidation result is informative because it fails, and the nature of that failure is instructive. In a continual learning protocol on 100 visual categories, geometric merging of nearby embeddings achieved 62.5\% compression but increased backward interference from $-0.100 \pm 0.003$ to $-0.394 \pm 0.034$ (Extended Data Fig.~\ref{fig:ed_consolidation}). The vector averaging fallacy is now visible in quantitative terms: centroid merging erases the fine angular structure that separates semantically adjacent memories, collapsing distinct traces into a blurred centroid that confuses retrieval. This result has a dual implication. For artificial systems, it is a direct warning: any vector database that implements deduplication or compression via centroid merging will predictably degrade retrieval fidelity. For biological memory, it rules out simple geometric merging as a model of hippocampal--neocortical consolidation and indicates that the brain's transfer mechanism must involve importance weighting, replay-guided refinement\cite{kumaran2012}, or reconsolidation-like updating\cite{nader2003}; the hippocampus maintains pattern-separated representations while the neocortex extracts statistical regularities through interleaved replay\cite{mcclelland1995}.

We note several important scope limitations and quantitative discrepancies that indicate where the geometric account is complete and where biology adds something the geometry alone does not capture. All experiments used English-language data; whether the geometric correspondences hold across typologically diverse languages would test the generality of these results directly. The TOT operationalisation (correct item ranked 2--20) is loose relative to the phenomenological criterion in humans and may partly explain the rate discrepancy. The parameter dependence of the spacing result means it is more contingent than the interference and DRM findings. The forgetting exponent ($b = 0.460 \pm 0.183$) matches the human value but with higher variance across seeds (range: 0.342--0.823) than human data shows, suggesting that biological systems may have stabilising mechanisms, such as attentional gating or consolidation, that reduce variance without changing the mean. The spacing effect shows the correct ordering but with a wider dynamic range (long $= 0.994$ versus human ${\sim}0.65$), suggesting that biological noise and interference are better matched to produce intermediate retention values than our parameter choices. The TOT rate ($3.66\%$) exceeds the human baseline by ${\sim}2.4\times$, suggesting that the geometric definition of TOT states (correct item ranked 2--20) may be looser than the phenomenological criterion in humans. Topological analysis was limited to 1{,}000-point subsamples, suggesting that higher-order topological features at larger scales remain to be characterised. Each discrepancy points toward a specific biological constraint that narrows the parameter space beyond what geometry alone specifies. What remains to be tested is whether alternative decay functions, noise models, or retrieval rules produce qualitatively different conclusions, and whether the correspondences reported here extend to non-English data and to production retrieval systems with hybrid architectures.

The convergence of multiple human-like phenomena from a single geometric framework is stronger than analogy. Analogy would mean that these systems behave similarly. What we show is that they fail for the same reason: the mathematics of similarity-based retrieval in finite-dimensional spaces, operating under noise and competition, produces power-law forgetting, semantic confusability, and partial retrieval states whether the substrate is silicon or cortex. The rich phenomenology of human memory, long attributed to the complexity of biological mechanisms, may in substantial part reflect these geometric constraints. Biology determines where in parameter space a given system sits. Geometry determines what happens when it gets there. For the core phenomena examined here, the boundary between biological and artificial memory is thinner than previously assumed.

\section*{Methods}

\subsection*{Models and architecture}

All experiments used exclusively open-weight models: Qwen2.5-7B\cite{qwen2024} (Apache~2.0) for answer generation, all-MiniLM-L6-v2\cite{reimers2019} (Apache~2.0) for sentence embeddings in Phases~1--2, BAAI/bge-base-en-v1.5\cite{xiao2023} (MIT) for Phases~2--3, BAAI/bge-large-en-v1.5\cite{xiao2023} (MIT) for Phase~5, and openai/clip-vit-base-patch32\cite{radford2021} (MIT) for image embeddings. The core embedding concatenates a content vector from a frozen pre-trained encoder with a context vector (positional, temporal, or episodic) and projects through a trained linear layer with LayerNorm. Phase~1 uses a 768$\to$384 projection trained with InfoNCE loss ($\tau = 0.07$, AdamW, lr$=10^{-3}$, batch$=256$, 10 epochs). Phase~2 adds 64-dim temporal encoding (three sinusoidal scales: 1-day, 30-day, 365-day). Phase~4 trains modality-specific projections (${\sim}10$K parameters each) into a shared 512-dim space with symmetric InfoNCE loss. Experiments were conducted on a cluster of four A100 GPUs (Supplementary Table~S8).

\subsection*{Forgetting and interference experiments}

\begin{figure}[!ht]
    \centering
    \includegraphics[width=\textwidth]{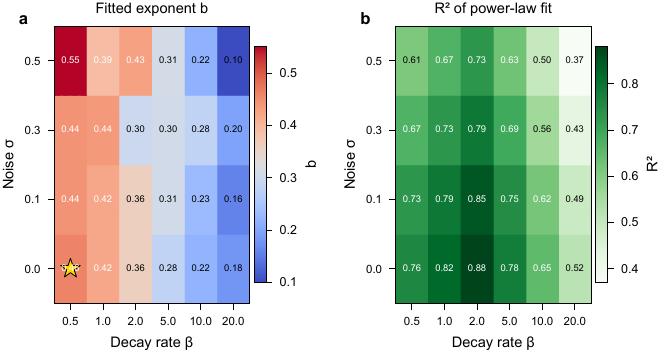}
    \caption{\textbf{Extended Data Fig.~3: Ebbinghaus parameter sweep.} \textbf{a},~Fitted forgetting exponent $b$ across the $\sigma \times \beta$ parameter grid. Contour line at $b = 0.5$ (human value). \textbf{b},~Goodness-of-fit $R^2$ for the same configurations.}
    \label{fig:ed_ebbinghaus}
\end{figure}

The Ebbinghaus simulation encoded 1{,}000 facts spanning 30 simulated days with 10{,}000 distractor sentences from TempLAMA. Retrieval scores were modulated by power-law decay: $S(t) = (1 + \beta t)^{-\psi}$, $\psi = 0.5$. This particular temporal kernel was chosen to match the functional form commonly used in the forgetting literature\cite{wixted1991}. A sweep over $\sigma \in \{0, 0.1, 0.3, 0.5\}$ and $\beta \in \{0.5, 1, 2, 5, 10, 20\}$ identified optimal configurations (Extended Data Fig.~\ref{fig:ed_ebbinghaus}). We note explicitly that the reported exponent sits within a tuned region of parameter space; the result is that the exponent is achievable under reasonable parameters, not that it emerges for all parameter choices.

\begin{figure}[!ht]
    \centering
    \includegraphics[width=\textwidth]{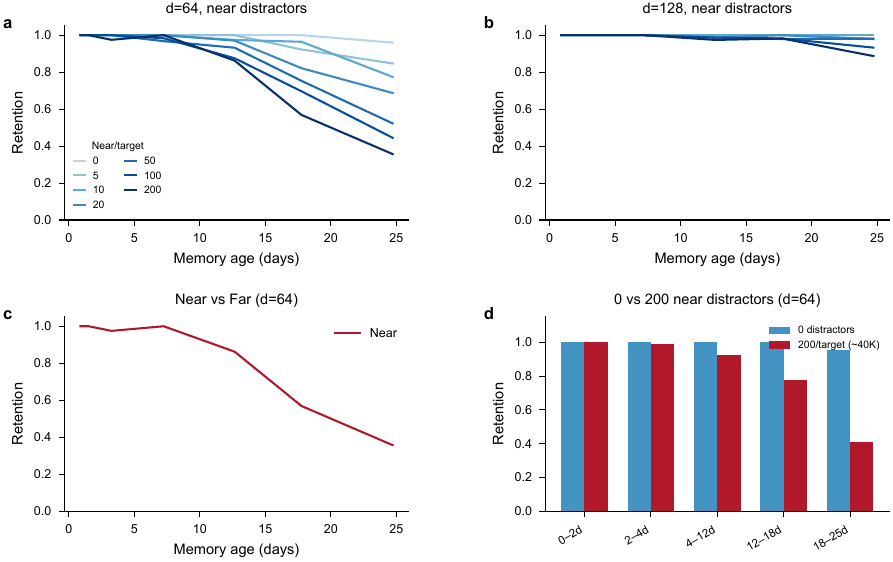}
    \caption{\textbf{Extended Data Fig.~4: Interference experiment full results.} \textbf{a},~All forgetting curves at $d = 64$ (near condition). \textbf{b},~Curves at $d = 128$ remain flat. \textbf{c},~Near versus far distractor comparison. \textbf{d},~Raw retention by age bin. $n = 5$ seeds.}
    \label{fig:ed_interference}
\end{figure}

The interference experiment selected 200 diverse Wikipedia articles (sampled to cover a range of topical domains; no deduplication filtering was applied beyond article-level selection), with one target sentence per article. Near distractors were same-article sentences; far distractors were cross-article sentences. Age-proportional noise: $\boldsymbol{\epsilon} = (\sigma \sqrt{a + 0.01}/\sqrt{d})\,\mathbf{z}$, $\mathbf{z} \sim \mathcal{N}(\mathbf{0}, \mathbf{I})$, $\sigma = 0.15$. This noise model is a synthetic stress model chosen because it introduces age-dependent degradation at a controlled rate; the core conclusion (that interference, not decay, drives the forgetting exponent) holds across the $\sigma$ and $\beta$ parameter ranges tested in the sweep (Extended Data Fig.~\ref{fig:ed_ebbinghaus}). BGE-large embeddings (1024-dim) were projected via PCA to $d \in \{64, 128, 256, 1024\}$. Near distractor counts: $\{0, 1\text{K}, 2\text{K}, 4\text{K}, 10\text{K}, 20\text{K}, 40\text{K}\}$; far counts: $\{0, 1\text{K}, 5\text{K}, 10\text{K}, 50\text{K}\}$ (Extended Data Fig.~\ref{fig:ed_interference}).

\subsection*{Emergent phenomena}

\begin{figure}[!ht]
    \centering
    \includegraphics[width=\textwidth]{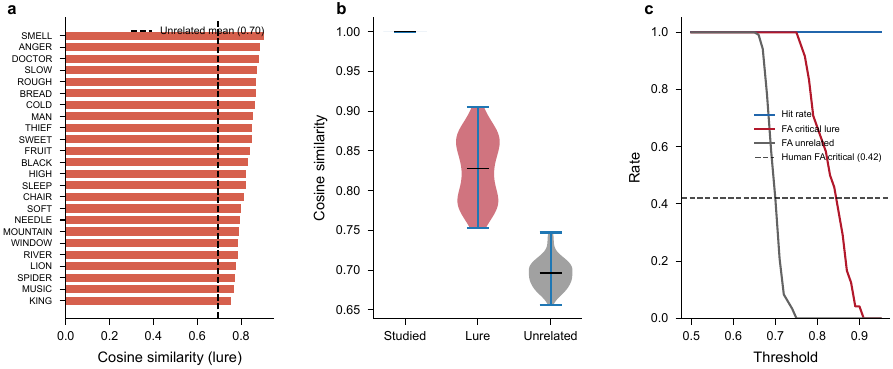}
    \caption{\textbf{Extended Data Fig.~7: DRM per-list analysis.} \textbf{a},~Lure cosine similarity across all 24 DRM word lists. \textbf{b},~Similarity distributions for studied words, critical lures, and unrelated words. \textbf{c},~Threshold operating curves for hit rate and false alarm rates. $n = 5$ seeds.}
    \label{fig:ed_drm}
\end{figure}

\textit{DRM false memory.} All 24 published word lists\cite{roediger1995} (15 studied words + 1 critical lure each) were encoded with BGE-large. Threshold sweep: $\theta \in [0.50, 0.95]$, step 0.01. No parameters were adjusted to match the human false alarm rate; the threshold $\theta = 0.82$ was selected on the independent criterion of zero unrelated false alarms (an operating convention rather than a theoretically privileged threshold), and the critical-lure rate at this threshold was observed, not optimised. Per-list results in Extended Data Fig.~\ref{fig:ed_drm}.

\begin{figure}[!ht]
    \centering
    \includegraphics[width=\textwidth]{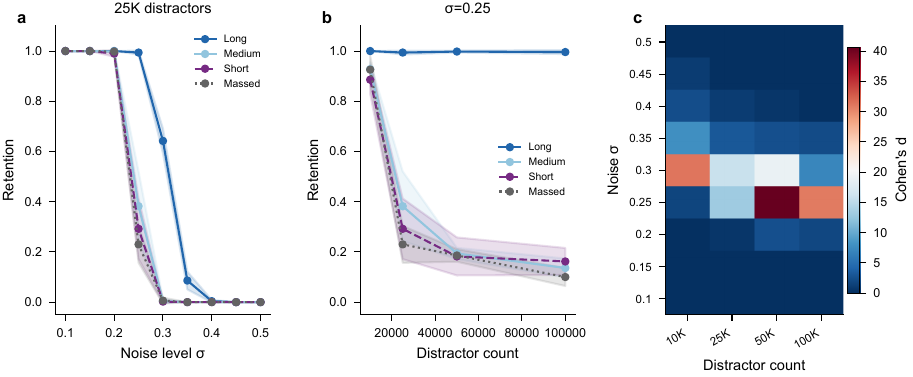}
    \caption{\textbf{Extended Data Fig.~8: Spacing sweep full results.} \textbf{a},~Retention by noise level at 25K distractors. \textbf{b},~Retention by distractor count at $\sigma = 0.25$. \textbf{c},~Effect size (Cohen's $d$, long versus massed) across the full $\sigma \times$ distractor grid. $n = 5$ seeds.}
    \label{fig:ed_spacing}
\end{figure}

\textit{Spacing effect.} 100 facts, 3 repetitions at 4 spacing conditions (massed: 0--2 min; short: 0--2 h; medium: 0--2 d; long: 0--2 w), tested at $t = 30$ d. Age-proportional noise: $\tilde{\mathbf{e}} = \text{normalize}(\mathbf{e} + \sigma\sqrt{a+0.01}/\sqrt{d}\cdot\mathbf{z})$. Full sweep in Extended Data Fig.~\ref{fig:ed_spacing}.

\begin{figure}[!ht]
    \centering
    \includegraphics[width=\textwidth]{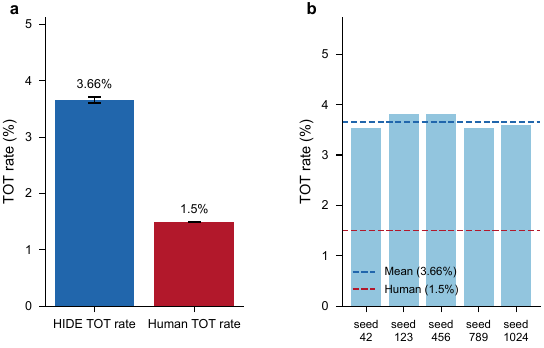}
    \caption{\textbf{Extended Data Fig.~9: Tip-of-tongue analysis.} TOT rate ($3.66 \pm 0.13\%$) compared to the human baseline (${\sim}1.5\%$). The qualitative phenomenon (correct items known but not immediately retrieved) emerges from retrieval competition, though the rate exceeds the human value by a factor of ${\sim}2.4$, suggesting the geometric operationalisation (rank 2--20) is looser than the human phenomenological criterion. $n = 5$ seeds.}
    \label{fig:ed_tot}
\end{figure}

\textit{Tip-of-tongue.} Embeddings projected to 128-dim via PCA; query noise $\sigma_q = 1.2/\sqrt{128}$. TOT: correct rank $>1$ and $\leq 20$ with top-1 similarity $> 0.5$. Analysis in Extended Data Fig.~\ref{fig:ed_tot}.

\textit{Topology.} Persistent homology via Rips complex (\texttt{gudhi}\cite{gudhi2014}) on 1{,}000-point subsamples of 10{,}000 Wikipedia embeddings. Betti numbers at $\epsilon \in \{0.3, 0.5, 0.7, 0.9, 1.0, 1.2, 1.5, 2.0, 2.5, 3.0\}$, max dimension~2.

\subsection*{Effective dimensionality analysis}

The participation ratio $d_\text{eff} = (\sum_i \lambda_i)^2 / \sum_i \lambda_i^2$, where $\lambda_i$ are the PCA eigenvalues of the embedding matrix, provides a continuous measure of the effective number of dimensions occupied by the data. We computed $d_\text{eff}$, $d_{95}$ (number of components for 95\% cumulative variance), and $d_{99}$ on 10{,}000 Wikipedia sentences for each of three models: MiniLM-L6-v2 (384-dim), BGE-base-en-v1.5 (768-dim), and BGE-large-en-v1.5 (1{,}024-dim). The MiniLM interference experiment used the same protocol as the main interference experiment (200 targets, age-proportional noise $\sigma = 0.15$, near distractor counts $\{0, 5, 10, 20, 50, 100, 200\}$) but on native MiniLM embeddings without PCA projection.

\subsection*{Datasets and statistical analysis}

All datasets publicly available: bAbI QA en-10k (BSD), TempLAMA (MIT), CIFAR-100 (BSD), COCO Captions 2017 (CC BY 4.0), Flickr30k (Research), DRM word lists\cite{roediger1995} (public domain), Wikipedia (${\sim}200{,}000$ sentences from 2{,}345 articles). All experiments: 5 seeds $[42, 123, 456, 789, 1024]$, mean $\pm$ std, bootstrap 95\% CI from 10{,}000 resamples. Cohen's $d$ for human comparisons. Full hyperparameters in Supplementary Table~S1.

\subsection*{Reproducibility}

\begin{figure}[!ht]
    \centering
    \includegraphics[width=\textwidth]{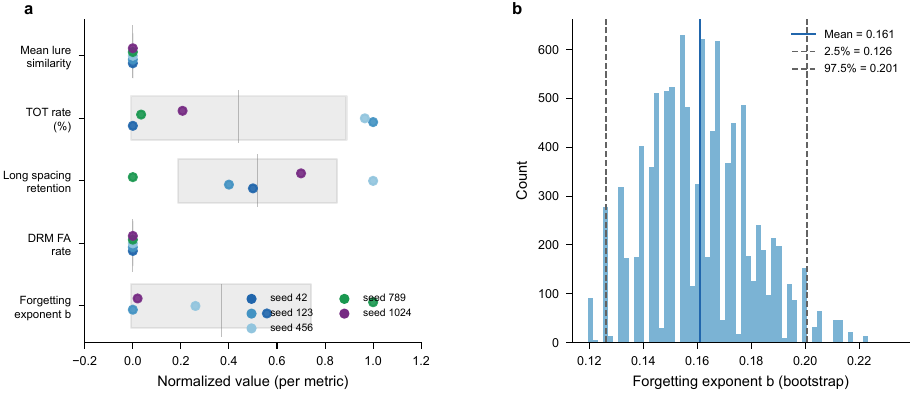}
    \caption{\textbf{Extended Data Fig.~10: Reproducibility across seeds.} \textbf{a},~Key metrics across all five random seeds, showing consistency of results. \textbf{b},~Bootstrap distribution of the forgetting exponent $b$ with 95\% confidence interval.}
    \label{fig:ed_reproducibility}
\end{figure}

All hyperparameters stored in YAML configuration files. Results saved as JSON and CSV. A \texttt{run\_all.py} script reproduces all experiments. Code and configurations available in the accompanying repository. Extended Data Fig.~\ref{fig:ed_reproducibility} shows consistency of key metrics across all five random seeds.

\section*{Author Contributions}
A.G.\ conceived the project, developed the theoretical framework and designed the experiments. A.G, A.S., S.R.B., S.B., and N.N.\ contributed to implementation, experimental execution and manuscript preparation.

\section*{Data Availability}

All datasets are publicly available through HuggingFace or standard repositories. DRM word lists are reproduced from published public-domain sources\cite{roediger1995}. No proprietary or synthetic data were used.

\section*{Code Availability}

All code, configuration files, and reproduction instructions are available in the accompanying repository under an open-source licence at \url{https://github.com/Dynamis-Labs/hide-project}.

\section*{Acknowledgements}

Code generation assisted by Claude (Anthropic).

\section*{Competing Interests}
The authors have financial interests in Dynamis Labs, Inc.


\clearpage
\section*{Extended Data}

\clearpage
\section*{Supplementary Information}

\subsection*{Table S1: Hyperparameters}

\begin{table}[!ht]
\centering
\caption{Hyperparameters across all phases. All values stored in YAML configuration files.}
\label{tab:hyperparams}
\small
\begin{tabular}{@{}llll@{}}
\toprule
\textbf{Phase} & \textbf{Parameter} & \textbf{Value} & \textbf{Notes} \\
\midrule
\multirow{7}{*}{1} & Embedding model & MiniLM-L6-v2 & 384-dim \\
 & Positional encoding dim & 384 & Sinusoidal \\
 & ContextProjector & $768 \to 384$ & Linear + LayerNorm \\
 & InfoNCE temperature & 0.07 & Fixed \\
 & Learning rate & $1 \times 10^{-3}$ & AdamW \\
 & Weight decay & $1 \times 10^{-4}$ & --- \\
 & Batch size / Epochs & 256 / 10 & 10\% validation split \\
\midrule
\multirow{6}{*}{2} & Temporal encoding dim & 64 & Three sinusoidal scales \\
 & Temporal scales & 1d, 30d, 365d & Fine, medium, coarse \\
 & Decay methods & Power law & $(1+\beta t)^{-\psi}$, $\psi = 0.5$ \\
 & Distractors (Ebbinghaus) & 10{,}000 & TempLAMA sentences \\
 & Interference: near distractors & 0--40{,}000 & Same-article, 7 levels \\
 & Interference: far distractors & 0--50{,}000 & Cross-article, 5 levels \\
 & Interference: PCA dims & $\{64, 128, 256, 1024\}$ & BGE-large base \\
 & Beta sweep & $\{0.5, 1, 2, 5, 10, 20\}$ & Noise $\sigma \in \{0, 0.1, 0.3, 0.5\}$ \\
\midrule
\multirow{5}{*}{3} & CLIP encoder & ViT-B/32 & 512-dim, frozen \\
 & HDBSCAN min\_cluster\_size & 10 & Gentle consolidation \\
 & Age-based selection & min\_age threshold & Protects recent memories \\
 & Consolidation frequency & Every 1{,}000 memories & Selective merging \\
 & Replay & 100 random old memories & Per consolidation cycle \\
\midrule
\multirow{5}{*}{4} & Shared dim & 512 & --- \\
 & Text projection & $384 \to 512$ & ${\sim}10$K params \\
 & Image projection & $512 \to 512$ & ${\sim}10$K params \\
 & InfoNCE temperature & 0.07 (learnable) & Symmetric \\
 & Epochs / Early stopping & 20 / patience 5 & Validation R@1 \\
\midrule
\multirow{8}{*}{5} & Embedding model & BGE-large & 1024-dim \\
 & DRM word lists & 24 lists & 15 words + 1 lure each \\
 & DRM threshold sweep & $\theta \in [0.50, 0.95]$ & Step 0.01 \\
 & Spacing noise $\sigma$ & 0.25 (best graded) & Sweep: $\{0.1, 0.15, \ldots, 0.5\}$ \\
 & Spacing distractors & 25{,}000 (best graded) & Sweep: $\{10\text{K}, 25\text{K}, 50\text{K}, 100\text{K}\}$ \\
 & TOT PCA dim & 128 & From 1024-dim \\
 & TOT query noise & $\sigma_q = 1.2$ & Dim-normalized \\
 & Topology subsample & 1{,}000 of 10{,}000 & Rips complex, max dim 2 \\
\bottomrule
\end{tabular}
\end{table}

\subsection*{Table S2: Dataset Statistics}

\begin{table}[!ht]
\centering
\caption{Datasets used across all phases. All publicly available with permissive licences.}
\label{tab:datasets}
\small
\begin{tabular}{@{}lllll@{}}
\toprule
\textbf{Dataset} & \textbf{Phase} & \textbf{Split} & \textbf{Size} & \textbf{Licence} \\
\midrule
bAbI QA (en-10k) & 1 & Train / Test & 10K / 1K per task & BSD \\
TempLAMA & 2 & Full & Variable & MIT \\
CIFAR-100 & 3 & Train / Test & 50K / 10K & BSD \\
COCO Captions 2017 & 4 & Train / Val & 118K / 5K & CC BY 4.0 \\
Flickr30k & 4 & Test & 1K & Research \\
DRM word lists & 5 & --- & 24 lists $\times$ 16 words & Public domain \\
Wikipedia (20231101.en) & 2, 5 & Streaming & 200{,}000 sentences (2{,}345 articles) & CC BY-SA \\
\bottomrule
\end{tabular}
\end{table}

\subsection*{Table S3: Phase 1 Full Results}

\begin{table}[!ht]
\centering
\caption{bAbI accuracy (mean $\pm$ std, 5 seeds) across tasks and methods.}
\label{tab:phase1_full}
\small
\begin{tabular}{@{}lccccc@{}}
\toprule
\textbf{Method} & \textbf{Task 1} & \textbf{Task 2} & \textbf{Task 3} & \textbf{Task 4} & \textbf{Task 5} \\
\midrule
Contextual retrieval & $0.597 \pm 0.009$ & $0.341 \pm 0.014$ & $0.281 \pm 0.012$ & $0.365 \pm 0.006$ & $0.793 \pm 0.008$ \\
No memory & $0.000$ & $0.000$ & $0.006$ & $0.011$ & $0.000$ \\
Full context & $0.773$ & $0.487$ & $0.267$ & $0.305$ & $0.840$ \\
Random retrieval & $0.531 \pm 0.008$ & $0.317 \pm 0.011$ & $0.229 \pm 0.013$ & $0.360 \pm 0.006$ & $0.620 \pm 0.018$ \\
\midrule
\textbf{Overall} & \multicolumn{5}{c}{Contextual: $0.475 \pm 0.007$; No memory: $0.003$; Random: $0.411 \pm 0.004$} \\
\bottomrule
\end{tabular}
\end{table}

\subsection*{Table S4: Interference Experiment}

\begin{table}[!ht]
\centering
\caption{Fitted power-law exponent $b$ and retention by distractor count, condition, and dimensionality (mean, 95\% CI, 5 seeds). MiniLM rows show native 384-dim without PCA projection, confirming that effective dimensionality, not nominal dimensionality, governs interference.}
\label{tab:interference}
\small
\begin{tabular}{@{}rllccc@{}}
\toprule
\textbf{Dim} & \textbf{Cond.} & \textbf{Distractors} & \textbf{$b$} & \textbf{95\% CI} & \textbf{Retention} \\
\midrule
\multirow{7}{*}{64} & near & 0 & 0.009 & [0.006, 0.012] & 0.990 \\
 & near & 1{,}000 & 0.038 & [0.023, 0.062] & 0.961 \\
 & near & 4{,}000 & 0.073 & [0.057, 0.095] & 0.919 \\
 & near & 10{,}000 & 0.105 & [0.087, 0.124] & 0.878 \\
 & near & 40{,}000 & 0.161 & [0.126, 0.200] & 0.814 \\
 & far & 10{,}000 & 0.078 & [0.054, 0.116] & 0.917 \\
 & far & 50{,}000 & 0.132 & [0.105, 0.156] & 0.850 \\
\midrule
128 & near & 40{,}000 & 0.020 & [0.015, 0.025] & 0.978 \\
256 & near & 40{,}000 & 0.003 & [0.002, 0.005] & 0.995 \\
1024 & near & 40{,}000 & 0.002 & [0.001, 0.004] & 0.998 \\
\midrule
\multicolumn{6}{c}{\textit{MiniLM-L6-v2 at native $d = 384$ ($d_\text{eff} \approx 16$), no PCA}} \\
\midrule
384$^*$ & near & 0 & 0.678 & [0.583, 0.789] & 0.268 \\
384$^*$ & near & 1{,}000 & 1.069 & [0.521, 1.640] & 0.047 \\
384$^*$ & near & 4{,}000 & 0.000 & --- & 0.000 \\
384$^*$ & near & 40{,}000 & 0.000 & --- & 0.000 \\
\bottomrule
\end{tabular}
\end{table}

\subsection*{Table S4b: Effective Dimensionality of Embedding Models}

\begin{table}[!ht]
\centering
\caption{Effective dimensionality analysis. Despite 25$\times$--65$\times$ differences in nominal dimensionality, all models concentrate variance in ${\sim}16$ effective dimensions. $N = 10{,}000$ Wikipedia sentences per seed, 5 seeds.}
\label{tab:eff_dim}
\small
\begin{tabular}{@{}lccccl@{}}
\toprule
\textbf{Model} & $d_\text{nom}$ & $d_\text{eff}$ & $d_{95}$ & $d_{99}$ & $d_\text{eff}/d_\text{nom}$ \\
\midrule
MiniLM-L6-v2 & 384 & $15.7 \pm 0.0$ & $17.0 \pm 0.0$ & $19.0 \pm 0.0$ & 0.041 \\
BGE-base-en-v1.5 & 768 & $16.6 \pm 0.1$ & $18.0 \pm 0.0$ & $19.0 \pm 0.0$ & 0.022 \\
BGE-large-en-v1.5 & 1{,}024 & $16.3 \pm 0.1$ & $17.6 \pm 0.5$ & $19.0 \pm 0.0$ & 0.016 \\
\bottomrule
\end{tabular}
\end{table}

\subsection*{Table S5: Consolidation Results}

\begin{table}[!ht]
\centering
\caption{Consolidation conditions on CIFAR-100 (mean $\pm$ std, 5 seeds).}
\label{tab:phase3_full}
\small
\begin{tabular}{@{}lccc@{}}
\toprule
\textbf{Condition} & \textbf{Backward Transfer} & \textbf{Compression} & \textbf{Notes} \\
\midrule
No consolidation & $-0.100 \pm 0.003$ & $1.000$ & Baseline \\
Gentle consolidation & $-0.394 \pm 0.034$ & $0.375 \pm 0.036$ & Age-based \\
Full (gentle + replay) & $-0.398 \pm 0.030$ & $0.432 \pm 0.056$ & + replay \\
\bottomrule
\end{tabular}
\end{table}

\subsection*{Table S6: Cross-Modal Retrieval}

\begin{table}[!ht]
\centering
\caption{Cross-modal retrieval on Flickr30k (mean $\pm$ std, 5 seeds).}
\label{tab:phase4_full}
\small
\begin{tabular}{@{}lcccccc@{}}
\toprule
& \multicolumn{3}{c}{\textbf{Image $\to$ Text}} & \multicolumn{3}{c}{\textbf{Text $\to$ Image}} \\
\cmidrule(lr){2-4} \cmidrule(lr){5-7}
\textbf{Method} & R@1 & R@5 & R@10 & R@1 & R@5 & R@10 \\
\midrule
Contextual & $0.203$ & $0.470$ & $0.602$ & $0.231$ & $0.492$ & $0.620$ \\
Random & $0.001$ & $0.007$ & $0.018$ & $0.002$ & $0.012$ & $0.021$ \\
\bottomrule
\end{tabular}
\end{table}

\subsection*{Table S7: Emergent Phenomena}

\begin{table}[!ht]
\centering
\caption{Emergent phenomena summary (mean $\pm$ std, 5 seeds).}
\label{tab:phase5_full}
\small
\begin{tabular}{@{}lccl@{}}
\toprule
\textbf{Phenomenon} & \textbf{Observed} & \textbf{Human} & \textbf{Notes} \\
\midrule
DRM FA (critical lure) & $0.583$ & ${\sim}0.55$ & Unbaked \\
Spacing: long & $0.994 \pm 0.008$ & $0.65$ & $\sigma = 0.25$, 25K \\
Spacing: massed & $0.230 \pm 0.073$ & $0.30$ & $\sigma = 0.25$ \\
TOT rate & $3.66 \pm 0.13\%$ & $1.5\%$ & PCA 128-dim \\
$H_1$ peak & $534 \pm 24$ & --- & At $\epsilon = 1.0$ \\
\bottomrule
\end{tabular}
\end{table}

\subsection*{Table S8: Compute Resources}

\begin{table}[!ht]
\centering
\caption{Computational resources. All experiments on 4$\times$ NVIDIA A100 GPUs.}
\label{tab:compute}
\small
\begin{tabular}{@{}llll@{}}
\toprule
\textbf{Phase} & \textbf{GPUs} & \textbf{Operations} & \textbf{Time} \\
\midrule
1 & 0, 1 & Embedding + generation & ${\sim}2$h \\
2 & 0, 1 & Temporal + Ebbinghaus & ${\sim}1$h \\
2 (interference) & 1 & Distractor sweep & ${\sim}3$h \\
3 & 1, 2 & CLIP + HDBSCAN & ${\sim}3$h \\
4 & 1, 2 & Projection + retrieval & ${\sim}2$h \\
5 & 1--3 & DRM + spacing + topology & ${\sim}2$h \\
\bottomrule
\end{tabular}
\end{table}

\subsection*{Table S9: DRM Per-List Results}

\begin{table}[!ht]
\centering
\caption{Per-list DRM cosine similarities and false alarm rates at $\theta = 0.82$ (mean across 5 seeds).}
\label{tab:drm_perlist}
\small
\begin{tabular}{@{}lcccl@{}}
\toprule
\textbf{List} & \textbf{Studied sim} & \textbf{Lure sim} & \textbf{Unrelated sim} & \textbf{FA rate} \\
\midrule
SLEEP & 1.000 & 0.821 & 0.672 & 0.583 \\
NEEDLE & 1.000 & 0.830 & 0.699 & 0.583 \\
ROUGH & 1.000 & 0.819 & 0.693 & 0.583 \\
SWEET & 1.000 & 0.835 & 0.700 & 0.583 \\
\multicolumn{5}{c}{\textit{(Full table with all 24 lists in Extended Data Fig.~\ref{fig:ed_drm})}} \\
\bottomrule
\end{tabular}
\end{table}

\end{document}